\documentclass[showpacs,pre]{revtex4}
\usepackage{graphicx}
\begin{document}
\title{Soliton turbulences in the complex Ginzburg-Landau equation}
\author{Hidetsugu Sakaguchi\\
Department of Applied Science for Electronics and Materials, \\
Interdisciplinary Graduate School of Engineering Sciences, Kyushu
University, \\
Kasuga, Fukuoka 816-8580, Japan
}
\begin{abstract}
We study spatio-temporal chaos in the complex Ginzburg-Landau equation in parameter regions of weak amplification and viscosity. Turbulent states involving many soliton-like pulses appear in the parameter range, because the complex Ginzburg-Landau equation is close to the nonlinear Schr\"odinger equation. We find that the distributions of amplitude and wavenumber of pulses depend only on the ratio of the two parameters of the amplification and the viscosity. This implies that a one-parameter family of soliton turbulence states characterized by different distributions of the soliton parameters exists continuously around the completely integrable system.
\end{abstract}
\pacs{05.45.Jn, 05.45.Yv, 42.65.Tg}
\maketitle
  Dissipative structures and chaos have been studied in nonlinear-nonequilibrium systems. Various kinds of spatio-temporal chaos or weak turbulences were found in Rayleigh-B\'enard convections, electro-hydrodynamic convections of liquid crystals and chemical systems~\cite{rf:1,rf:2}. The Kuramoto-Sivashinsky equation is one of the simplest model which exhibits spatio-temporal chaos, and the statistical properties of the spatio-temporal chaos have been intensively studied~\cite{rf:3,rf:4,rf:5,rf:6}.  The complex Ginzburg-Landau equation has been also intensively studied as a model of spatio-temporal chaos~\cite{rf:7,rf:8,rf:9}. The Kuramoto-Sivashinsky equation appears as a phase equation for the complex Ginzburg-Landau equation near the onset of the phase instability. The energy spectrum of the spatio-temporal chaos in the Kuramoto-Sivashinsky equation exhibits almost flat in a large scale (small wavenumber regime), which has an analogy with the equi-partition law of energy in thermal equilibrium states. On the other hand, in hydrodynamics turbulences, which are described by the Navier-Stokes equation, the energy spectrum has a singular form called the Kolmogorov spectrum and the strong turbulences exhibit multi-fractal intermittency~\cite{rf:10}. These behaviors are closely related to the singular behavior of the Euler equation, which appears in the limit of no viscosity of the Navier-Stokes equation.  In the limit of no viscosity and no amplification, the complex Ginzburg-Landau equation is reduced to the nonlinear Schr\"odinger equation, which is well known as a completely integral system and a typical soliton equation. 
In this brief report, we study the statistical properties of the complex Ginzburg-Landau equation in the regime of weak amplification and viscosity. The complex Ginzburg-Landau equation with weak amplification and viscosity appears naturally in some problems of nonlinear optics such as optical fibers with dissipation and external pumping~\cite{rf:11}. Kishiba et al. tried to explain the energy spectrum in the complex Ginzburg-Landau equation with weak amplification, saturation and viscosity~\cite{rf:12}. 
In the spatio-temporal chaos in the regime of weak amplification and viscosity, soliton-like pulses play an important role and it is rather different from the spatio-temporal chaos studied before in Refs.[8] and [9]. We will show its peculiar statistical properties of the soliton turbulences and discuss a relation with the behavior in the nonlinear Schr\"odinger equation.

Our model equation has the form

\begin{equation}
i\frac{\partial \phi}{\partial t}+\frac{1}{2}\frac{\partial^2\phi}{\partial x^2}+|\phi|^2\phi=i\left (\epsilon_1\phi+\epsilon_2\frac{\partial^2 \phi}{\partial x^2}\right ),
\end{equation}
where $\phi$ is a complex variable, $\epsilon_1$ and $\epsilon_2$ are  respectively parameters of amplification and viscosity. We consider the case of $\epsilon_1>0$ and $\epsilon_2>0$ in this paper.
In the limit of $\epsilon_1=\epsilon_2=0$, Eq.~(1) is reduced to the nonlinear Schr\"odinger equation and has a family of soliton solutions:
\begin{equation}
\phi=\frac{A e^{ik(x-vt)-i\omega t}}{\cosh\{A(x-vt)\}},
\end{equation}
where $A$ is the amplitude of the soliton, $k$ is the wavenumber, and $v=k$ is the velocity of the soliton. 
The amplitude and the wavenumber can be independently changed as soliton parameters. The nonlinear Schr\"odinger equation is a completely integrable system for infinite length ($L=\infty$). Then, the nonlinear Schr\"odinger equation has an infinite number of invariants including the total norm $N=\int_{-\infty}^{\infty}|\phi|^2dx$ and the total momentum $P=\int_{-\infty}^{\infty}(i\partial \phi^*/\partial x\phi-i\partial \phi/\partial x\phi^*)/2dx$, and so on.  If $\epsilon_1>0$ or $\epsilon_2>0$, Eq.~(1) becomes a dissipative system, and the invariants of motion disappear. 

We have performed numerical simulations of Eq.~(1) with the split-step Fourier method. The timestep is $\Delta t=0.001$, and periodic boundary conditions are imposed. The system size $L$ is a control parameter. Even for very small parameters of $\epsilon_1$ and $\epsilon_2$, the chaotic behavior appears for large system. 
Regular behavior appears in a small system. Figure 1(a) displays a temporal evolution of $|\phi|$ at $\epsilon_1=\epsilon_2=0.01$ and $L=6$. Two pulses appear from a uniform state $\phi=0$, because $\phi=0$ is an unstable solution. The peak positions do not change in time at $L=6$. However, the peak amplitude of the pulses is breathing in time. The breathing motions of the two pulses are out of phase as in seen in Figure 1(b). As $L$ is increased, the peak positions begin to move. An example of moving pulses is shown in Fig.~1(c) at $L=6.7$. The spatial motion of the two pulses is almost synchronized but the breathing motions are out of phase in time.   

\begin{figure}[tbp]
\includegraphics[width=13cm]{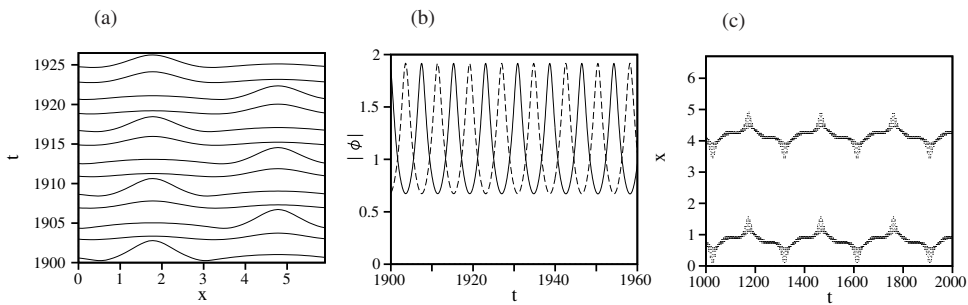}
\caption{(a) Time evolution of $|\phi|$ at $\epsilon_1=\epsilon_2=0.01$ and $L=6$. (b) Time evolution of peak amplitudes $|\phi(x)|$ at $\epsilon_1=\epsilon_2=0.01$ and $L=6$. The solid (dashed) line denotes the peak amplitude of the pulse located in $x<L/2$ ($x>L/2$). (c) Time evolution of peak positions of the two pulses for $L=6.7$ at $\epsilon_1=\epsilon_2=0.01$.
} \label{fig1}
\end{figure}
\begin{figure}[tbp]
\includegraphics[width=13cm]{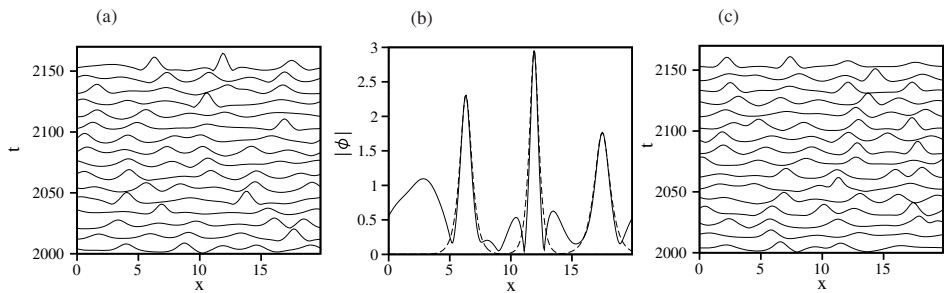}
\caption{(a) Time evolution of $|\phi|$ for $L=20$ at $\epsilon_1=\epsilon_2=0.01$. (b) Snapshot profile of $|\phi(x,t)|$ at $t=2150$. Dashed curves denote approximation by soliton-type solutions. (c)  Time evolution of $|\phi|$ for $L=20$ at $\epsilon_1=\epsilon_2=0.$. 
} \label{fig2}
\end{figure}
When the system size $L$ is further increased, more pulses are created and spatio-temporal chaos appears. Figure 2(a) displays chaotic time evolution of $|\phi(x,t)|$ for $L=20$ at $\epsilon_1=\epsilon_2=0.01$. 
Creation and annihilation of pulse structures occur, and radiation-like waves with small amplitudes also appear. Figure 2(b) is a snapshot profile of $|\phi(x,t)|$ at $t=2150$. Three dashed curves denote $2.3/\cosh\{2.3(x-6.32)\}, 2.95/\cosh\{2.95(x-11.9)\}$ and $1.76/\cosh\{1.76(x-17.47)\}$.  The pulse structures are well approximated by the soliton solutions Eq.~(2), because the parameters $\epsilon_1$ and $\epsilon_2$ are rather small. The spatiotemporal chaos might be therefore interpreted as soliton turbulence. Figure 2(c) displays time evolution of $|\phi(x,t)|$ for $t>2000$ in another numerical simulation, in which $\epsilon_1=\epsilon_2$ takes the same value 0.01 as in the case of Fig.~2(a) before $t=1900$, but $\epsilon_1=\epsilon_2=0$ are set to zero after $t=1900$. That is, $\phi$ obeys the nonlinear Schr\"odinger equation after $t=1900$. The initial condition at the fresh start time $t=1900$ is a state which has appeared as a result of the spatio-temporal chaos. 
The time evolution in Fig.~2(c) is not chaotic,  because the nonlinear Schr\"odinger equation is a completely integrable system. However, the time evolution of $|\phi|$ in Fig.~2(c) looks similar to Fig.~2(a). 

To characterize statistical properties of the spatio-temporal chaos,
we have calculated the modulus $A=|\phi|$ at the local maxmum points in the profile $|\phi(x,t)|$, which is interpreted as an amplitude of soliton, and  a local wavenumber of soliton: $k=(-i/2)(\phi_x\phi^*-\phi_x^*\phi)/|\phi|^2$ at $t=5n$ ($n$ is an integer). And we have constructed a distribution $P(A)$ and $P(k)$ of $A$ and $k$ in a larger system $L=80$. The parameters $\epsilon_1$ and $\epsilon_2$ are changed as $\epsilon_1=\epsilon_2=0.01,0.001,0.0001$ and 0, and $\epsilon_1=\epsilon_2/2=0.01,0.001,0.0001$ and 0. The two initial conditions in the case of  $\epsilon_1=\epsilon_2=0$ are the two final states which were numerically obtained for the parameters $\epsilon_1=\epsilon_2=0.0001$ and $\epsilon_1=\epsilon_2/2=0.0001$.
The numerical simulation was performed until $t=30000$ and the data between $t=20000$ and $30000$ were used to construct the distributions. 
\begin{figure}[tbp]
\includegraphics[width=9cm]{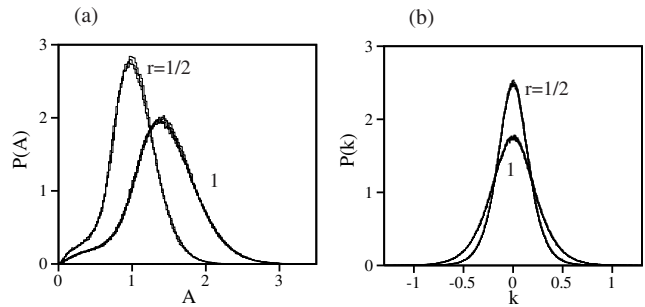}
\caption{(a) Distributions $P(A)$ of amplitude of pulses. (b) Distributions $P(k)$ of wavenumber of pulses. Time evolution of $|\phi|$ for $L=20$ at $\epsilon_1=\epsilon_2=0.01$. Eight distributions for eight parameter sets of $\epsilon_1=\epsilon_2=0.01,0.001,0.0001$ and 0, and $\epsilon_1=\epsilon_2/2=0.01,0.001,0.0001$ and 0 are plotted. The ratio $r=\epsilon_1/\epsilon_2$ for the parameter sets is 1 or 1/2, and the value is denoted near the distributions.
} \label{fig3}
\end{figure}
\begin{figure}[tbp]
\includegraphics[width=13cm]{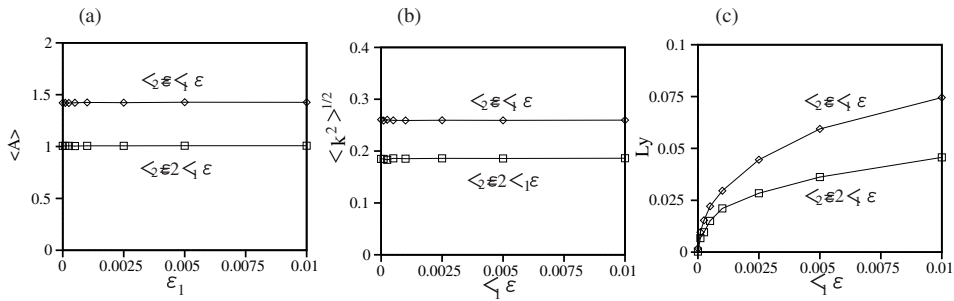}
\caption{(a) Average value of $A$ as a function of $\epsilon_1$ for $\epsilon_2=\epsilon_1$ and $\epsilon_2=2\epsilon_1$. (b) Root mean square of wavenumber $k$ as a function of $\epsilon_1$ for $\epsilon_2=\epsilon_1$ and $\epsilon_2=2\epsilon_1$. (c) The first Lyapunov exponent as a function of $\epsilon_1$ for $\epsilon_2=\epsilon_1$ and $\epsilon_2=2\epsilon_1$.
} \label{fig4}
\end{figure}
Figures 3(a) and (b) display the distribution $P(A)$ and $P(k)$ for the eight parameter sets. The distributions $P(A)$ and $P(k)$ overlap very well, when the ratio $r=\epsilon_1/\epsilon_2$ takes the same value 1 or 1/2. The overlap is seen even in the limit of $\epsilon_1=\epsilon_2=0$. This is consistent with the result seen in Figs.~2(a) and (c). That is, the statistical properties of the spatio-temporal chaos and the integrable system are almost the same, if the initial condition for the nonlinear Schr\"odinger equation is chosen as a state in the spatio-temporal chaos.  On the other hand, the distributions for $\epsilon_1/\epsilon_2=1$ and $1/2$ are definitely different. 
We have calculated the average value of $A$ and the root mean square of $k$ for $\epsilon_1=\epsilon_2=0.01,0.005,0.0025,0.001,0.0005,0.00025,0.0001,0$ and $\epsilon_1=\epsilon_2/2=0.01,0.005,0.0025,0.001,0.0005,0.00025,0.0001,0$.  
The results are shown in Figs.~4(a) and (b). Almost horizontal lines imply that the distributions of $P(A)$ and $P(k)$ depend only on the ratio $r=\epsilon_1/\epsilon_2$, when $\epsilon_1$ is sufficiently small.  We have also calculated the first Lyapunov exponent which characterizes the spatio-temporal chaos.
Small deviation $\delta \phi$ from $\phi(x,t)$ obeys a linearized equation of Eq.~(1):
\begin{equation}
i\frac{\partial (\delta \phi)}{\partial t}+\frac{1}{2}\frac{\partial^2(\delta\phi)}{\partial x^2}+\{2|\phi|^2(\delta\phi)+\phi^2(\delta\phi)^*\}=i\left (\epsilon_1(\delta\phi)+\epsilon_2\frac{\partial^2 (\delta\phi)}{\partial x^2}\right ).
\end{equation}
The first Lyapunov exponent was calculated as the average value of the linear growth rate of the quantity $[\int_0^L|\delta\phi|^2dx]^{1/2}$.
Figure 4(c) displays the first Lyapunov exponent for the same parameter sets as in Figs.~4(a) and (b). The first Lyapunov exponent increases from 0 as a function of $\epsilon_1$, which implies that the spatio-temporal chaos becomes stronger as $\epsilon_1$ is increased. The first Lyapunov exponent is naturally 0 for $\epsilon_1=\epsilon_2=0$. These behaviors seem to be strange, but they are not paradoxical, because the irregularity is not generated from a regular initial condition owing to the non-positive Lyapunov exponent, but the regularity is neither created from an irregular initial condition owing to the non-negative Lyapunov exponent in the nonlinear Schr\"odinger equation. 
\begin{figure}[tbp]
\includegraphics[width=9cm]{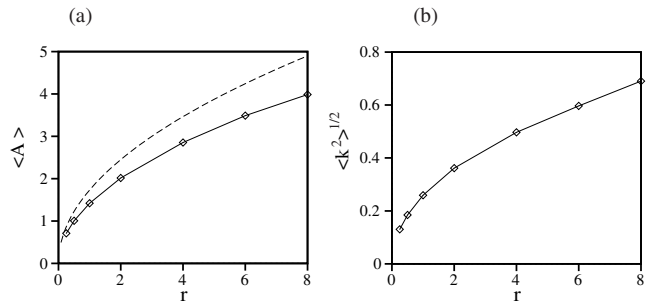}
\caption{(a) Average value of $A$ as a function of the ration $r=\epsilon_1/\epsilon_2$ for $\epsilon_2=0.001$. The dashed curve is $(3r)^{1/2}$. (b) Root mean square of wavenumber $k$ as a function of the ratio $r=\epsilon_1/\epsilon_2$ for $\epsilon_2=0.001$.  
} \label{fig5}
\end{figure}

 The distributions $P(A)$ and $P(k)$ depend on the ratio $r=\epsilon_1/\epsilon_2$ continuously. Figures 5(a) and 5(b) display the average value of $A$ and the root mean square of $k$ as a function of the ratio $r=\epsilon_1/\epsilon_2$ for a fixed value of $\epsilon_2=0.001$.
The average value of $A$ and the root mean square of $k$ is a increasing function of the ratio $r$. The total norm $N=\int_0^L|\phi|^2dx$ obeys
\begin{equation}
\frac{dN}{dt}=\int_0^L[(\partial \phi/\partial t)\phi^*+\phi(\partial \phi/\partial t)^*]dx=\int_0^L(2\epsilon_1|\phi|^2-2\epsilon_2|\partial \phi/\partial x|^2)dx.
\end{equation}
If $\phi(x,t)$ is approximated by a sum of solitons as $\phi(x)\sim\sum_iA_i\exp\{ik_i(x-x_i)\}{\rm sech}\{A(x-x_i)\}$, Eq.~(4) is approximated as
\begin{equation}
\frac{dN}{dt}=2\sum_i(\epsilon_1A_i-\epsilon_2A_i^3/3-\epsilon_2k_i^2 A_i).
\end{equation}
If the total norm is assumed to be constant in time, the distances $|x_i-x_j|$ between two solitons are large, $\langle A^3\rangle$ is approximated by $\langle A\rangle^3$ and $\langle k^2\rangle$ is neglected, $\langle A\rangle$ is estimated as $\langle A\rangle\sim (3\epsilon_1/\epsilon_2)^{1/2}=(3r)^{1/2}$, which is a rough estimate of the pulse amplitude and denoted by the dashed curve in Fig.~5(a). We have not succeeded in estimating the root mean square of $k$, yet. 

To summarize, we have found that soliton turbulence appears when $\epsilon_1$ and $\epsilon_2$ are sufficiently small in the complex Ginzburg-Landau equation (1). The soliton turbulence is a chaotic attractor of the complex Ginzburg-Landau equation, so the time evolution leads to the chaotic attractor from almost all initial conditions. The soliton turbulence is characterized by definite distributions of $P(A)$ and $P(k)$, and the distributions are determined by the ratio $r=\epsilon_1/\epsilon_2$. Even if $\epsilon_1$ is decreased to zero with a fixed ratio $r=\epsilon_1/\epsilon_2$, the distributions $P(A)$ and $P(k)$ take almost the same form. On the other hand, the Lyapunov exponent is decreased to zero, when $\epsilon_1$ is deacreased to zero. 
In the limit of $\epsilon_1=0$ and $\epsilon_2=0$, the complex Ginzburg-Landau equation is reduced to the nonlinear Schr\"odinger equation. In the nonlinear Schr\"odinger equation, there is no attractor and no ergodicity, and the time evolution is completely determined by the initial conditions. That is, a one-parameter family of soliton turbulence states corresponding to the different ratio $r=\epsilon_1/\epsilon_2$, which are characterized by different distributions of $A$ and $k$, exists around the completely integrable system. This is a unique relation between the weak spatio-temporal chaos exhibited by the complex Ginzburg-Landau equation and the completely integrable dynamics exhibited by the nonlinear Schr\"odinger equation.


\begin{thebibliography}{99}
\bibitem{rf:1}
P.~Manneville {\it Disspipative Structures and Weak Turbulence} (Academic Press, ,Boston, 1990.)
\bibitem{rf:2}
T.~Bohr, M.~H.~Jensen, G.~Paladin and A.~Vulpiani, {\it Dynamical Systems Approach to Turbulence} (Cambridge University Press, Cambridge, 1998).
\bibitem{rf:3}
Y.~Kuramoto, {Chemical Oscillations, Waves and Turbulence} (Springer-Verlag, Berlin, 1984).
\bibitem{rf:4} S.~Zaleski, Physica D {\bf 34}, 427 (1989).
\bibitem{rf:5} K.~Sneppen, J.~Krug, M.~H.~Jensen, C.~Jayaprakash, and T.~Bohr, Phys. Rev. A {\bf 46} R7351 (1992).
\bibitem{rf:6} H.~Sakaguchi, Phys. Rev. E {\bf 62}, 8817 (2000). 
\bibitem{rf:7} I.~S.~Aranson and L.~Kramer, Rev. Mod. Phys. {\bf 74},99 (2002).
\bibitem{rf:8} H.~Sakaguchi, Prog. Theor. Phys. {\bf 89}, 1123 (1993).
\bibitem{rf:9} D.~A.~Egolf and H.~S.~Greenside, Phys. Rev. Lett. {\bf 74}, 1751 (1995).
\bibitem{rf:10} U.~Frisch, {\it Turbulence} (Cambridge University Press, Cambridge, 1995).
\bibitem{rf:11} G.~P.~Agrawal, {\it Nonlinear Fiber Optics} (Academic Press, San Diego,2001).
\bibitem{rf:12} S.~Kishiba, S.~Toh, T.~Kawahara, Physica D {\bf 54}, 43 (1991).
\end{thebibliography}
\end{document}